\documentclass[aps,prd,preprint]{revtex4}

\usepackage{graphicx}
\usepackage{psfrag}

\begin{document}
\title{BCVEGPY2.0: A upgrade version of the generator
BCVEGPY with an addendum about hadroproduction of the $P$-wave
$B_c$ states}
\author{Chao-Hsi Chang$^{1,2}$ \footnote{email: zhangzx@itp.ac.cn},
Jian-Xiong Wang$^3$ \footnote{email:jxwang@mail.ihep.ac.cn} and
Xing-Gang Wu$^{3}$\footnote{email: wuxg@mail.ihep.ac.cn}}
\address{$^1$CCAST (World Laboratory), P.O.Box 8730, Beijing
100080, P.R. China.\footnote{Not correspondence address.}\\
$^2$Institute of Theoretical Physics, Chinese Academy of Sciences,
P.O.Box 2735, Beijing 100080, P.R. China.\\
$^3$Institute of High Energy Physics, Chinese Academy of Sciences,
P.O.Box 918(4), Beijing 100049, P.R. China.}

\begin{abstract}
The generator BCVEGPY is upgraded by adding the hadroproduction of
the $P$-wave excited $B_c$ states (denoted by $B_{cJ,L=1}^*$ or by
$h_{B_c}$ and $\chi_{B_c}$) and by improving some features of the
original one as well. We denote it as BCVEGPY2.0. The $P$-wave
production is also calculated by taking only the dominant
gluon-gluon fusion mechanism (with the subprocess $gg\to
B_{cJ,L=1}^*+\bar{c}+b$ being dominated) into account as that for
$S$-wave. In order to make the addendum piece of the upgraded
generator as compact as possible so as to increase its efficiency,
we manipulate the amplitude as compact as possible with FDC (a
software for generating Feynman diagrams and the algebra
amplitudes, and for manipulating algebra formulae analytically
etc) and certain simplification techniques. The correctness of the
program is tested by checking the gauge invariance of the
amplitude and by comparing the numerical results with the existent
ones in the literature carefully.


\end{abstract}
\maketitle


\noindent{\bf PROGRAM SUMMARY}\\

\noindent{\it Title of program} : BCVEGPY\\

\noindent{\it Version}: 2.0 (December, 2004)\\

\noindent{\it Program obtained from} : CPC Program Library or
Institute of Theoretical Physics, Chinese Academy of Sciences,
Beijing, P.R. China:
{$www.itp.ac.cn/\,\widetilde{}\;zhangzx/BCVEGPY2.0/$}.\\

\noindent{\it Reference to original program} : BCVEGPY1.0\\

\noindent{\it Reference in CPC} : Comput. Phys. Commun. {\bf 159},
192(2004)\\

\noindent{\it Does the new version supersede the old program} :
Yes\\

\noindent{\it Computer}: Any computer with FORTRAN 77 compiler.
The program has been tested on HP-SC45 Sigma-X parallel computer,
Linux PCs and Windows PCs with Virsual Fortran.\\

\noindent{\it Operating systems} : UNIX, Linux and Windows.\\

\noindent{\it Programming language used} : FORTRAN 77/90.\\

\noindent{\it Memory required to execute with typical data} :
About
2.0 MB.\\

\noindent{\it No. of bytes in distributed program, (including
PYTHIA6.2)} : About 1.1 MB.\\

\noindent{\it Distribution format} : Compressed tar file.\\


\noindent{\it Nature of physical problem} : Hadronic production of
$B_c$ meson and its $P$-wave excited states.\\

\noindent{\it Method of solution} : The code with option can
generate weighted and un-weighted events. For jet hadronization,
an interface to PYTHIA is provided.\\

\noindent{\it Restrictions on the complexity of the problem} : The
hadronic production of $c\bar{b}$-quarkonium in $S$-wave and
$P$-wave states via the mechanism of gluon-gluon fusion are given
by the so-called 'complete calculation' approach. The less
important contributions from the other mechanisms have not
been included.\\

\noindent{\it Typical running time} : It depends on which option
one runs to match PYTHIA when generating the $B_c$ events.
Typically, For the hadronic production of the $S$-wave
$c\bar{b}$-quarkonium, if IDWTUP=$1$, then it takes about 20 hour
on a 1.8 GHz Intel P4-processor machine to generate 1000 events;
however if IDWTUP=$3$, to generate $10^6$ events, it takes about
40 minutes only. For the hadronic production of the $P$-wave
$c\bar{b}$-quarkonium, the time will be almost two times longer
than the case of the $S$-wave quarkonium.\\

\noindent{\bf LONG WRITE-UP}

\section{Introduction}

Recently, we have completed the hadronic production of the
$P$-wave states of $B_c$ \cite{cww,cqww}, that has not been
included in the existent version BCVEGPY1.0 \cite{bcvegpy1} yet,
so we upgrade the generator here. The studies in
Refs.\cite{cww,cqww} involve not only the contributions from the
color-singlet components $|(c\bar{b})_{\bf 1}(^1P_1)\rangle$ and
$|(c\bar{b})_{\bf 1}(^3P_J)\rangle \, (J=1,2,3)$ but also those
from the color-octet components $|(c\bar b)_{\bf 8}(^{1}S_{0})
g\rangle$ and $|(c\bar b)_{\bf 8}(^{3}S_{1}) g\rangle$
\cite{nrqcd} to the $P$-wave production. Especially, it is found
that the $P$-wave production at Tevatron and LHC can be so big
even as that in the same order of the magnitude for the $S$-wave
production in certain kinematics regions. Therefore, to compare
with the original generator, in addition to some improvements on
the program for the $S$-wave $B_c$ (and $B_c^*$) production, the
main fresh feature of the upgraded version BCVEGPY2.0 is that it
can generate the hadroproduction of the $P$-wave excited $B_c$
states (denoted by $B_{cJ,L=1}^*$ or by $h_{B_c}$, $\chi^0_{B_c}$,
$\chi^1_{B_c}$ and $\chi^2_{B_c}$) as well. At such high energies
of Tevatron and LHC, the gluon-gluon fusion mechanism i.e. that
via the subprocess $gg\to B_{cJ,L=1}^*+\bar{c}+b$, is also
dominant for the $P$-wave production \cite{cww,cqww}, hence in the
addendum we only take the gluon-gluon fusion mechanism into
account.

Therefore, in fact, here we just need to rewrite the program used
in Refs.\cite{cww,cqww} into the format according to the
environment PYTHIA \cite{pythia} as the original version
BCVEGPY1.0 properly.

Besides that the hadronic production of the $P$-wave $B_c$ states
is added, for convenience and increasing efficiency, some
improvements in BCVEGPY2.0 are made. They can be summarized as:
i). an improved way to record the $B_c$ events is available, i.e.,
all the possible information, such as various distributions of the
$B_c$ events with proper kinematic cuts, can be recorded just by
running the program once; ii). an improved way is available to
generate the useful sampling importance function just by running
VEGAS program \cite{gpl} carefully; iii). an `extra switch' is
added, i.e. whether one switches on or off to adopt the newly
improved parton distribution functions (PDFs) (other than the
inner PDFs provided by PYTHIA itself) to the simulation, so that
one can apply the latest version of the PDFs provided by the
several groups such as CTEQ \cite{6lcteq}, GRV \cite{98lgrv}, MRS
\cite{2001lmrst} conveniently; iv). the color flow of the
processes is re-written (as an important issue for the new
version), so as to well-match that in PYTHIA.

In BCVEGPY1.0, the color flow is treated according to the usual
fundamental-representation decomposition \cite{cf0}. Since the
color flow for a given hard scattering subprocess in PYTHIA is
treated in the way as that in Ref.\cite{cf2}, i.e., the color
fundamental-representation decomposition is replaced by the
color-flow decomposition equivalently under the large $N_c$ limit.
Such a color-flow decomposition has a more intuitive physical
interpretation and practically it is easier to deal with than the
usual fundamental-representation decomposition. Hence in order to
well-match with PYTHIA and to be easy to deal with, especially,
with the shower Monte Carlo for QCD jets, the replacement to treat
the color flow as that in Ref.\cite{cf2} for the relevant
sub-processes is adopted in  BCVEGPY2.0.

In order to increase the efficiency and to make the $P$-wave
production amplitude as compact as possible, we manipulate the
amplitude with the help of some techniques and the software FDC (a
software for generating Feynman diagrams and the algebra
amplitudes, and for manipulating algebra formulae analytically etc
\cite{fdc}). To guarantee the correctness of the program, as
Refs.\cite{cww,cqww}, we have made numerical checks on the gauge
invariance of the amplitude and numerical comparisons between the
present obtained results and those in the literature carefully.

The paper is organized: following Introduction, in Section II we
outline the new features of BCVEGPY2.0 in detail, and explain how
to use and test (check) the program. In Sec.III, we present some
discussions on the generator and make a summary.

\section{The program: BCVEGPY2.0}

Since in hadron collisions at high energies, the gluon-gluon
fusion mechanism for the hadronic production of the $B_c$ meson is
dominant over the other mechanisms, thus in BCVEGPY1.0
\cite{bcvegpy1}, the hadronic production of the $S$-wave $B_c$ and
$B_c^*$ mesons via the gluon-gluon fusion mechanism has been
accomplished. While the present target for the new version
BCVEGPY2.0 now essentially is to add the hadronic production of
the $P$-wave $B_c$ states to the generator BCVEGPY. In BCVEGPY2.0,
for the production of the four $P$-wave $B_{cJ,L=1}^*$ ($h_{B_c}$,
$\chi^0_{B_c}$, $\chi^1_{B_c}$ and $\chi^2_{B_c}$) states, not
only the contributions from color-singlet components $|(c\bar
b)_{\bf 1}(^{1}P_{1}) \rangle$ and $|(c\bar b)_{\bf
1}(^{3}P_{J=0,1,2})\rangle$ but also those from the color-octet
components $|(c\bar b)_{\bf 8}(^{1}S_{0}) g\rangle$ and $|(c\bar
b)_{\bf 8}(^{3}S_{1}) g\rangle$ are included. The formulation and
additional techniques in calculating the hadronic production of
$P$-wave states have been given in Refs.\cite{cww,cqww} with
explanations, hence for simplicity, we do not repeat them here.
The readers may refer to Refs.\cite{cww,cqww} for details if they
are interested in them.

To supersede the old version BCVEGPY1.0 by the new version
BCVEGPY2.0 conveniently, BCVEGPY2.0 is also written into a Fortran
package, and it can generate the $S$-wave and the $P$-wave $B_c$
states accordingly with proper options. In BCVEGPY2.0, for
comparison, the hadronic production of $S$-wave states
$B_c(^1S_0)$ and $B_c^*(^3S_1)$ via the less important light
quark-antiquark annihilation subprocess, $q+\bar{q}\rightarrow
B_c(B^*_c)+\bar{c}+b$, (the annihilation mechanism) has also been
included. Since BCVEGPY2.0 is also written in the format as PYTHIA
(including common block variables), thus all the functions of
PYTHIA can be utilized in connection with the use of BCVEGPY2.0.

\subsection{New features in BCVEGPY2.0}

BCVEGPY2.0 is written in the same structure as that of BCVEGPY1.0,
with some improvements in addition to the $P$-wave production
being merged. It contains the following new features:

{$\bullet$} The amplitudes for the hadronic production of the
color-singlet corresponding to the four $P$-wave states,
$B_{cJ,L=1}^*$ or $^1P_1$ and $^3P_J$ (J=0,1,2), are included.
Four files to calculate the amplitudes for the $P$-wave states are
added: p1p1amp.for, pj0amp.for, pj1amp.for and pj2amp.for, which
are for the amplitudes of $^1P_1$ and $^3P_J$ (J=0,1,2)
respectively. While the necessary subroutines and functions for
calculating the square of the amplitudes for all the $P$-wave
states are put into the files psamp.for and lorentz.for. In fact,
the file lorentz.for includes basic subroutines, which are to
calculate the polarization vectors and the polarization tensor of
the heavy quarkonium, and to generate short notations for typical
expressions of useful contracts of the vectors and/or tensors so
as to make the program more compact. Three extra head files:
inclamp.f, inclcon.f and inclppp.f are included in the package,
which contain the frequently used common blocks for calculating
each hard scattering amplitude. Note here, some new techniques to
compute the $P$-wave amplitude \cite{cww}, such as to expand the
amplitude by `independent and elementary fermion strings'
analytically with the help of the FDC program \cite{fdc} etc., are
used instead of the original helicity ones that are used in
computing the $S$-wave amplitude.

{$\bullet$} To give full estimates of the $P$-wave production up
to $v^2$, we also take into account the contributions to the
hadronic $P$-wave production from the color-octet components
$|(c\bar b)_{\bf 8}(^{1}S_{0}) g\rangle$ and $|(c\bar b)_{\bf
8}(^{3}S_{1}) g\rangle$. In the cases of the hadronic production
$gg\rightarrow (c\bar{b})_{\bf 1} +b+\bar{c}$ with
$(c\bar{b})_{\bf 1}$ in color-singlet, $|(c\bar b)_{\bf
1}(^{1}S_{0})\rangle$ or $|(c\bar b)_{\bf 1}(^{3}S_{1})\rangle$,
there are only three independent color factors (i.e. three
independent color flows \cite{cww,cf}). While in the cases of the
$P$-wave hadronic production due to the color-octet components,
such as $gg\rightarrow (c\bar{b})_{\bf 8} +b+\bar{c}$ with
$(c\bar{b})_{\bf 8}$ in color-octet, there are totally ten
independent color factors \cite{cqww}. We shall show in more
detail about how to deal with the color-flows in the next
subsection. The hard scattering amplitudes corresponding to the
color-octet components can be directly read from the Eqs.(4-8) in
Ref.\cite{bcvegpy1} but the color-singlet matrix elements (or the
wave function at origin) are replaced by the color-octet matrix
elements $\langle 0|\chi_b^\dag T^d\psi_c \ (a_H^\dag
a_H)\psi_c^\dag T^d\chi_b |0\rangle$ and $\langle 0|\chi_b^\dag
\sigma^i T^d \psi_c\ (a_H^\dag a_H)\psi_c^\dag \sigma^i T^d
\chi_b|0\rangle$ accordingly. We can approximately relate the
color-octet matrix elements to the color-singlet one by NRQCD
scaling rules \cite{nrqcd} as the method described in
Ref.\cite{wccf}. Applying the heavy-quark spin symmetry, the above
two production matrix elements are related by \cite{nrqcd}
\begin{eqnarray}
\langle 0|\chi_b^\dag \sigma^i T^d \psi_c\ (a_H^\dag
a_H)\psi_c^\dag \sigma^i T^d \chi_b|0\rangle= (2 J +
1)\cdot\langle 0|\chi_b^\dag T^d\psi_c \ (a_H^\dag a_H)\psi_c^\dag
T^d\chi_b |0\rangle[1+{\cal O}(v^2)],
\end{eqnarray}
where $J$ is the total angular momentum of the hadron state. More
specifically, based on the velocity scale rule\cite{nrqcd}, we
estimate
\begin{eqnarray}
\langle 0|\chi_b^\dag T^d\psi_c \ (a_H^\dag a_H)\psi_c^\dag
T^d\chi_b |0\rangle &\simeq& \Delta_S(v)^2 \cdot \langle
0|\chi_b^\dag \psi_c (a_H^\dag a_H)
\psi_c^\dag \chi_b |0\rangle\nonumber\\
&\simeq& \Delta_S(v)^2\cdot \left|\langle 0 |\chi_b^\dag
\psi_c|B_c(^1S_0)\rangle\right|^2.\left[1+{\cal
O}(v^4)\right]\label{deltas}
\end{eqnarray}
where the second equation comes from the vacuum-saturation
approximation. $v$ is the relative velocity between the constitute
quarks in the bound states. $\Delta_S(v)$ is of the order $v^2$ or
so, and one can take it to be within the region of 0.10 to 0.30,
which is in consistent with the identification:
$\Delta_S(v)\sim\alpha_s(Mv)$.

{$\bullet$} The $S$-wave ($^1S_0$ and $^3S_1$) hadronic production
via the less important mechanism, i.e. the light quark-antiquark
annihilation mechanism with the subprocess $q+\bar{q}\rightarrow
B_c(B^*_c)+\bar{c}+b$, is also included in BCVEGPY2.0. In the
light quark-antiquark annihilation mechanism, the technique as
that in the gluon-gluon fusion one \cite{bcvegpy1} is applied, but
it is much more simple (only seven Feynman diagrams are needed to
be considered), so we do not present any detail for the
calculation here. Numerical results of the calculation show that
its contributions to the production are smaller than those from
the gluon-gluon fusion ($\sim 1\%$) \cite{prod1,changwu}.

{$\bullet$} For convenience, in order to record more data at one
run which may interest us, we add 24 extra files to record the
information of the generated events, and then all these output
files are put in a subdirectory named data. For example, the
generated files, pt005y.dat, pt020y.dat, pt035y.dat, pt050y.dat
and pt100y.dat are used to record the rapidity distributions with
different $p_t$ cuts, i.e. $p_{tcut}=5, 20, 35, 50, 100GeV$,
respectively. The user may record other interested information in
a similar way.

{$\bullet$} With the option IVEGASOPEN=1, one can use the VEGAS
program \cite{gpl} to achieve an important sampling-function,
which may be used to increase the efficiency of the Monte Carlo
simulation \cite{pythia}. Otherwise one may set IVEGASOPEN=0.
Furthermore, in order to improve the precision of the existed
important sampling-function obtained by previous VEGAS run, we add
one more input parameter IVEGGRADE, i.e., with IVEGGRADE=1, one
can use the existed sampling important function recorded in the
file grade.dat to generate a more precise important
sampling-function and record it in a new file newgrade.dat.
Otherwise one should set IVEGGRADE=0.

{$\bullet$} PYTHIA itself provides sixteen parton distribution
functions (PDFs). It is known that, in fact, the PDFs are obtained
through global fitting of the experimental data and then involuted
to the requested characteristic scale $Q^2$ by a standard way of
perturbative Quantum Chromodynamics (pQCD). Several groups, CTEQ
\cite{6lcteq}, GRV \cite{98lgrv} and MRS \cite{2001lmrst} etc,
devote themselves to offer accurate PDFs to the world and upgrade
them continually with fresh experimental data being available.
BCVEGPY2.0 provides a switch (IOUTPDF) to determine which type of
PDF one uses: if (IOUTPDF=1), then one can use one of the latest
version of PDFs from the above three groups with a proper value
for the parameter (IPDFNUM) to do the hadronic production;
otherwise, with (IOUTPDF=0), one can directly use a certain
existent PDF (by setting the value of the PYTHIA parameter
MSTP(51)) provided in PYTHIA to do the hadronic production. For
convenience, we have downloaded all the source files for the
latest version of the leading order (LO) PDFs of those three
groups, i.e. CTEQ6L, GRV98L and MRST2001L, and saved them into a
single fortran file outerpdf.for. The three corresponding data
files (cteq6l.tbl, grv98lo.grid and lo2002.dat), which are
necessary when running the file outerpdf.for for these three PDFs,
have been also added in the package. In fact, it is shown by
numerical calculations that the uncertainties for the estimates
due to the different PDFs can reach up to the level $\sim 10\%$ in
both cases for the $S$-wave and $P$-wave hadronic production
\cite{changwu,cww}.

{$\bullet$} For convenience, we have written an additional file
parameter.for to set the initial values of the parameters shown in
TAB.I. One may fix the value of `{\bf IREADDATA}' in parameter.for
as zero (`{\bf IREADDATA=0}') and then one returns back the
original way (that in version 1.0) to set the input parameters by
reading the parameter values from the date file: totput.dat. In
practice, if one want to generate a large number of results with
various input parameters, we would like to suggest him/her to
adopt the way of reading the input data directly from the date
files accordingly. In this way, only the input files need to be
changed and all the source files (*.for) need not to be changed,
so we can save a lot of time on making the source files and
furthermore, all the works can be written in a single executable
file and be executed automatically.

\begin{table}
\label{input0}
\begin{center}
\caption{The parameter values in the sequential order in the
totput.dat file.}
\begin{tabular}{|l|} \hline   PMBC
$\;\;\;$PMB$\;\;\;$ PMC$\;\;\;$ FBC\\ PTCUT $\;\;\;$ETACUT$\;\;\;$
ECM $\;\;\;$IBCSTATE$\;\;\;$
IGENERATE$\;\;\;$ IVEGASOPEN\\
NUMBER $\;\;\;$ NITMX \\
NQ2 $\;\;\;$NPDFU$\;\;\;$ NEV \\
ISHOWER$\;\;\;$ MSTP(51)$\;\;\;$
IDWTUP$\;\;\;$ MSTU(111)$\;\;\;$ PARU(111)\\
ISUBONLY$\;\;\;$ SUBENERGY$\;\;\;$ IGRADE\\
INUMEVNT$\;\;\;$ IVEGGRADE$\;\;\;$ IQQBAR$\;\;\;$ IQCODE\\
IOUTPDF$\;\;\;$ IPDFNUM$\;\;\;$ IOCTET$\;\;\;$  COEOCT\\
\hline
\end{tabular}
\end{center}
\end{table}

In summary, in the new package BCVEGPY2.0, there are twenty one
files totally, which includes thirteen main source files:
bcvegpy.for, foursets.for, genevnt.for, lorentz.for, outerpdf.for,
p1p1amp.for, parameter.for, pj0amp.for, pj1amp.for, pj2amp.for,
psamp.for, py6208.for (PYTHIA version 6.208), ssamp.for; three
date files for the corresponding three PDFs: lo2002.dat,
cteq6l.tbl, grv98lo.grid; three head files: inclamp.f, inclcon.f,
inclppp.f; one input data file: totput.dat and finally a simple
help file: readme.dat.

With all these new features in mind, the sequential order and the
format of the parameters in the input data file, totput.dat, now
changes to Table.I. Some new parameters specified in the input
file are:

\noindent $\bullet$ {\bf FBC=:} is different from the meaning in
BCVEGPY1.0, and now it's value is the radial wave function $R(0)$
for $S$-wave production or the first derivative of the radial wave
function $R'(0)$ for $P$-wave production. In our calculations, we
take FBC=$1.24GeV^{3/2}$, which corresponds to
$\psi(0)=1/\sqrt{4\pi}R(0)=0.35GeV^{3/2}$ for $S$-wave, and
FBC=$0.448GeV^{5/2}$, which corresponds to
$\psi'(0)=\sqrt{3/4\pi}R'(0)=0.219GeV^{5/2}$ for $P$-wave.

\noindent $\bullet $ {\bf IBCSTATE=:} indicates the $S$- and
$P$-states of $B_c$ meson: IBCSTATE=$1$, for $B_c[^{1}S_{0}]$;
IBCSTATE=$2$, for $B_c^*[^{3}S_{1}]$; IBCSTATE=$3,4,\cdots,6$, for
the four $P$-wave states $B_c^*[^{1}P_{1}]$, $B_c^*[^{3}P_{0}]$,
$B_c^*[^{3}P_{1}]$ and $B_c^*[^{3}P_{2}]$ respectively.

\noindent $\bullet $ {\bf INUMEVNT=:} indicates whether to switch
on or off to keep the event number distributions other than the
differential distributions of $p_T$, $y$ and etc.. INUMEVNT=$1$
means to switch on and INUMEVNT=$0$ means to switch off.

\noindent $\bullet $ {\bf IVEGGRADE=:} indicates whether to switch
on or off to use the existed grade (importance sampling function)
before running VEGAS. IVEGGRADE=$1$ means to switch on and
IVEGGRADE=$0$ means to switch off. With this parameter, one can
improve the efficiency of the importance sampling function.

\noindent $\bullet $ {\bf IQQBAR=:} indicates whether to switch on
or off to use the less important light quark-antiquark
annihilation subprocess $q+\bar{q}\rightarrow
(c\bar{b})+\bar{c}+b$ to generate the $(c\bar{b})$-quarkonium
events, where $(c\bar{b})$-quarkonium is in $^1S_0$ or $^3S_1$
state. IQQBAR=$1$ means to switch on and IQQBAR=$0$ means to
switch off. When IQQBAR=$0$, it will use the default gluon-gluon
fusion subprocess to generate the $(c\bar{b})$-quarkonium events.

\noindent $\bullet $ {\bf IQCODE=:} indicates the species of the
light quark for the light quark-antiquark annihilation mechanism.
Under the option IQQBAR=$1$, IQCODE=$1$ means $u$ and $\bar{u}$,
IQCODE=$2$ means $d$ and $\bar{d}$, while IQCODE=$3$ means $s$ and
$\bar{s}$ used in the quark-antiquark annihilation subprocess.

\noindent $\bullet $ {\bf IOUTPDF=:} indicates whether to switch
on or off to use the three latest version of PDFs: CTEQ6L, GRV98L
and MRST2001L offered in the program for the hadronic production.
IOUTPDF=$1$ means to switch on, while IOUTPDF=$0$ means to switch
off and then the inner PDFs of PYTHIA is used to generate the
events.

\noindent $\bullet $ {\bf IPDFNUM=:} indicates which one of the
three latest version of PDFs: CTEQ6L, GRV98L and MRST2001L is used
for the hadronic production. It comes into operation only under
the option IOUTPDF=$1$. IPDFNUM=$100$ means to use GRV98L;
IPDFNUM=$200$ means to use MRST2001L; IPDFNUM=$300$ means to use
CTEQ6L.

\noindent $\bullet $ {\bf IOCTET=:} indicates whether to switch on
or off to use the subprocess $gg\rightarrow (c\bar{b})_{\bf 8}
+b+\bar{c}$ to generate the color-octet components $|(c\bar
b)_{\bf 8}(^{1}S_{0}) g\rangle$ and $|(c\bar b)_{\bf 8}(^{3}S_{1})
g\rangle$. IOCTET=$1$ means to switch on and IOCTET=$0$ means to
switch off.

\noindent $\bullet $ {\bf COEOCT=:} the value for $\Delta_S(v)$,
where $\Delta_S(v)$ is defined in Eq.(\ref{deltas}).

\subsection{On the color flow for the hardronic production of $B_c$ meson}

A string-based fragmentation scheme for the cross sections, such
as that in the Lund model and used in PYTHIA, is needed for
different color flows. In order to merge well with the shower
Monte Carlo used in PYTHIA, in BCVEGPY2.0 we adopt the color-flow
decomposition that has been introduced in Ref.\cite{cf2} to
analyze the production. In Ref.\cite{cf2}, it is demonstrated that
the color-ordered amplitudes (i.e. amplitudes pertain to the
corresponding color-flows), that appear in the color-flow
decomposition, are identical to those appear in the usual
fundamental-representation decomposition under the large $N_c$
limit. So practically, one can first use the color-flow
decomposition to analyze how many independent color-flows we need
for a concerned subprocess and then transform the results into the
usual fundamental-representation decomposition according to the
color-flow Feynman rules. In this way, we can read out the
color-ordered amplitudes easily from our original results in
Refs.\cite{bcvegpy1,cww,cqww}.

Since the less important light quark-antiquark annihilation
mechanics with the subprocess $q+\bar{q}\rightarrow
B_c(B^*_c)+\bar{c}+b$ has only two independent orthogonal
color-flows, so we do not repeat the results here. We show the
color-flows for the dominant gluon-gluon fusion subprocess in
detail as the follows. In BCVEGPY2.0, we need to deal with the
color-flows for the dominant gluon-gluon fusion subprocess
$gg\rightarrow (c\bar{b}) +b+\bar{c}$ in two cases: one is for
$(c\bar{b})$-quarkonium in a color-singlet state (i.e.
$B_c$,$B^*_c$ and $B_{cJ,L=1}^*$) and the other is for
$(c\bar{b})$-quarkonium in a color-octet state (i.e. $|(c\bar
b)_{\bf 8}(^{1}S_{0}) g\rangle$ and $|(c\bar b)_{\bf 8}(^{3}S_{1})
g\rangle$).

\begin{figure}
\centering \setlength{\unitlength}{1mm}
\begin{picture}(80,60)(30,30)
\put(-10,-35) {\includegraphics{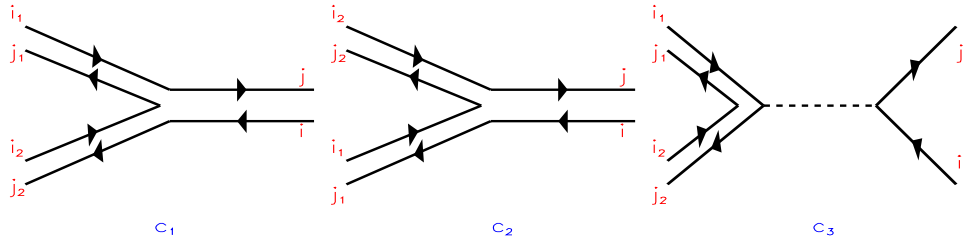}}
\end{picture}
\caption{ Color flow diagrams for the color-singlet production
based on the color-flow decomposition\cite{cf2}. Each pair of
indices $i_k$ and $j_k$ corresponds to an external gluon, i.e.
$k=1$ is for gluon-1 and $k=2$ for gluon-2. $i$ and $j$ are the
decomposed color indices for the outgoing $\bar{c}$ and $b$
respectively.}\label{cflow1}
\end{figure}

\noindent $\bullet $ The first case: $(c\bar{b})$-quarkonium in a
color-singlet state. In this case, we does not need to consider
the color structure of the $(c\bar{b})$-quarkonium and one may
find that there are only three types of independent color flows as
shown in Fig.\ref{cflow1}:
\begin{equation}
c_1=(\delta^j_{i_2}\delta^{j_2}_{i_1}\delta^{j_1}_{i}),\;
c_2=(\delta^j_{i_1}\delta^{j_1}_{i_2}\delta^{j_2}_{i}),\;
c_3=(\delta^j_{i}\delta^{j_2}_{i_1}\delta^{j_1}_{i_2}),
\end{equation}
where $(i_n,j_n)$ (n=1,2) are the color indices for the gluon-1
and gluon-2 respectively, $i$ and $j$ are color indices of the
outgoing $\bar{c}$ and $b$. In the conventions, the lower indices
$i$ and $i_n$ (n=1,2) transform under the anti
fundamental-representation of $SU(N_c)$ and upper indices $j$ and
$j_n$ (n=1,2) under the fundamental-representation. Note here that
the color bases are not normalized as done in Ref.\cite{cf2},
because we only need the relative probabilities among each color
flow. According to the color-flow Feynman rules \cite{cf2}, the
color-flow decomposition can be related to the
fundamental-representation through the following transformation:
\begin{equation}
c_1\rightarrow (T^aT^b)_{ij},\;\; c_2\rightarrow
(T^bT^a)_{ij},\;\; c_3\rightarrow (\delta_{ij}Tr[T^aT^b]).
\end{equation}
Equivalently, the total amplitude of the process can be written in
the fundamental-representation decomposition in the following way:
\begin{equation}\label{cos}
M=(T^aT^b)_{ij}M_1+ (T^bT^a)_{ij}M_2 +(\delta_{ij}Tr[T^aT^b])M_3,
\end{equation}
where $T^a$ and $T^b$ are the color matrices of the gluon-1 and
gluon-2; $i,j=1,2,3$ are color indices of the outgoing quarks
$\bar{c}$ and $b$ respectively. $M_i$ are the color-ordered
amplitudes respectively, which can be directly read out from
Refs.\cite{bcvegpy1,cww}.

\begin{figure}
\centering \setlength{\unitlength}{1mm}
\begin{picture}(80,60)(30,30)
\put(-5,-5) {\includegraphics{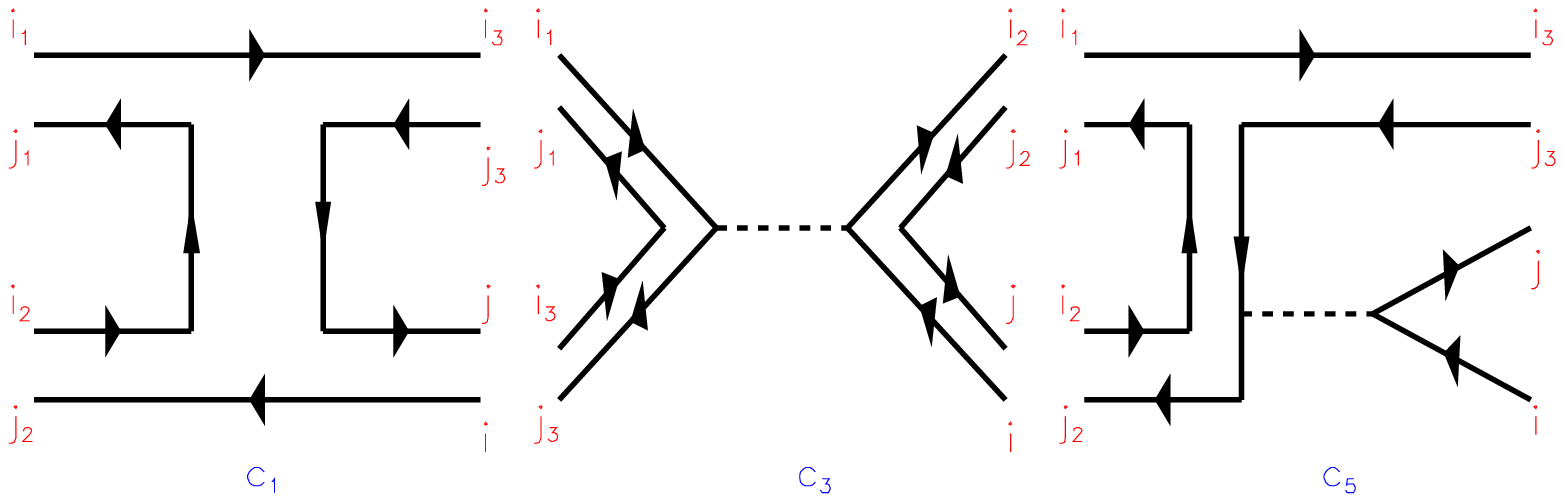}}
\put(15,-40) {\includegraphics{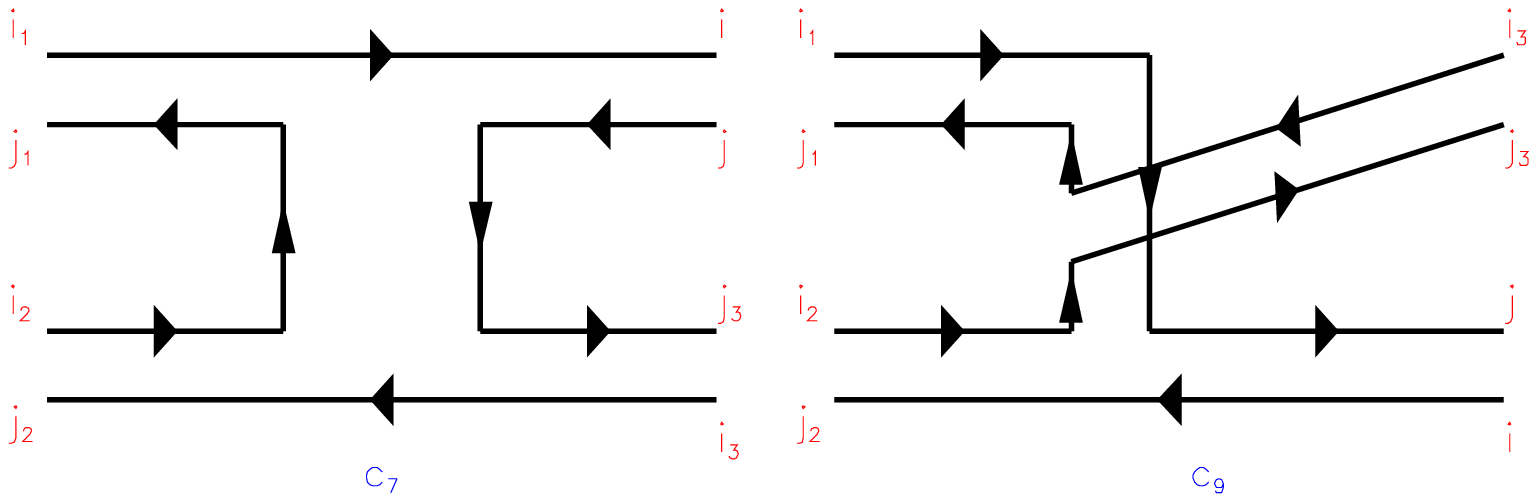}}
\end{picture}
\caption{ Color flow diagrams for the color-octet production based
on the color-flow decomposition\cite{cf2}. Each pair of indices
$i_k$ and $j_k$ corresponds to an external gluon, i.e. $k=1$ is
for gluon-1, $k=2$ for gluon-2 and $k=3$ for the color-octet
$(c\bar{b})$-quarkonium. $i$ and $j$ are the decomposed color
indices for the outgoing $\bar{c}$ and $b$ respectively. The
diagrams for $c_{n+1}$ $(n=1,2,\cdots,5)$ can be directly obtained
by gluon exchange.} \label{cflow2}
\end{figure}

\noindent $\bullet $ The second case: $(c\bar{b})$-quarkonium in
the color-octet state. In this case, we need to consider the color
structure of the $(c\bar{b})$-quarkonium itself, and one may find
that there are ten types of independent color flows totally as
shown in Fig\ref{cflow2}:
\begin{eqnarray}
&c_1=(\delta_i^{j_2}\delta_{i_2}^{j_1}\delta_{i_1}^{j_3}
\delta_{i_3}^{j})\,,\;\;\;
c_2=(\delta_i^{j_1}\delta_{i_1}^{j_2}\delta_{i_2}^{j_3}
\delta_{i_3}^{j})\,,\;\;\;
c_3=(\delta_{i_1}^{j_3}\delta_{i_3}^{j_1}\delta_{i}^{j_2}
\delta_{i_2}^{j})\,,\nonumber\\
&c_4=(\delta_{i_2}^{j_3}\delta_{i_3}^{j_2}\delta_{i}^{j_1}
\delta_{i_1}^{j})\,,\;\;\;
c_5=(\delta_{i}^{j}\delta_{i_2}^{j_1}\delta_{i_1}^{j_3}
\delta_{i_3}^{j_2})\,,\;\;\;
c_6=(\delta_{i}^{j}\delta_{i_1}^{j_2}\delta_{i_2}^{j_3}
\delta_{i_3}^{j_1})\,,\;\;\;c_7=(\delta_{i}^{j_3}\delta_{i_3}^{j_2}\delta_{i_2}^{j_1}
\delta_{i_1}^{j})\,,\nonumber\\
& c_8=(\delta_{i}^{j_3}\delta_{i_3}^{j_1}\delta_{i_1}^{j_2}
\delta_{i_2}^{j})\,,\;\;\;
c_9=(\delta_{i}^{j_1}\delta_{i_1}^{j_3}\delta_{i_3}^{j_2}
\delta_{i_2}^{j})\,,\;\;\;
c_{10}=(\delta_{i}^{j_2}\delta_{i_2}^{j_3}\delta_{i_3}^{j_1}
\delta_{i_1}^{j})\,,
\end{eqnarray}
where $i_3$ and $j_3$ are the decomposed color indices of $c$ and
$\bar{b}$ in the $(c\bar{b})$-quarkonium.  According to the
color-flow Feynman rules \cite{cf2}, the color-flow decomposition
can be related to the fundamental-representation through the
following `correspondences':
\begin{eqnarray}
&c_1\rightarrow (T^bT^aT^d)_{ij},\;\; c_2\rightarrow
(T^aT^bT^d)_{ij},\;\; c_3\rightarrow
(T^b_{ij}Tr[T^aT^d]),\nonumber\\
&c_4\rightarrow (T^a_{ij}Tr[T^bT^d]),\;\; c_5\rightarrow
(\delta_{ij}Tr[T^bT^aT^d]),\;\; c_6\rightarrow
(\delta_{ij}Tr[T^aT^bT^d]),\;\; c_7\rightarrow (T^dT^bT^a)_{ij},\nonumber\\
&c_8\rightarrow (T^dT^aT^b)_{ij},\;\; c_9\rightarrow
(T^aT^dT^b)_{ij},\;\;c_{10}\rightarrow (T^bT^dT^a)_{ij},
\end{eqnarray}
where $T^d$ stands for the color matrix of the color-octet
$(c\bar{b})$-quarkonium. Equivalently, the total amplitude of the
process can be written in the fundamental-representation
decomposition in the following,
\begin{eqnarray}\label{coo}
M&=&(T^bT^aT^d)_{ij}M_1+ (T^aT^bT^d)_{ij}M_2
+(T^b_{ij}Tr[T^aT^d])M_3+(T^a_{ij}Tr[T^bT^d])M_4 \nonumber\\
&+&(\delta_{ij}Tr[T^bT^aT^d])M_5+(\delta_{ij}Tr[T^aT^bT^d])M_6
+(T^dT^bT^a)_{ij}M_7+(T^dT^aT^b)_{ij}M_8\nonumber\\
&+& (T^aT^dT^b)_{ij}M_9+(T^bT^dT^a)_{ij}M_{10}.
\end{eqnarray}
$M_i$ are the corresponding color-ordered amplitudes, which can be
directly read out from the results in Refs.\cite{bcvegpy1,cqww}.

\noindent $\bullet $ In both cases, when squaring the amplitude
and summing over colors, the leading terms in $1/N_c$ obtained by
the square of each color flow. The cross terms between different
color flows are suppressed by a certain power of $N_c$ comparing
with the leading terms and under the large $N_c$ limit they can be
safely neglected. Therefore, we can obtain the probability for
each color flow, which are just $|M_i|^2$ accordingly, with
i=(1,2,3) in color-singlet case and i=(1,2,$\cdots$,10) in
color-octet case.

\subsection{Checks for the generator}

The amplitude for $S$-wave production has been checked well in the
original version already, thus the check for the whole Fortran
package BCVEGPY2.0 essentially is on the new contents in
comparison to the old version. As for BCVEGPY2.0, it is checked by
examining the gauge invariance of the amplitude for each
production, $S$-wave and $P$-wave of the $B_c$ meson,
individually, i.e. the amplitude vanishes when the polarization
vector of an initial gluon is substituted by the momentum vector
of this gluon. Numerically we find that the gauge invariance is
guaranteed at the computer ability (double precision).

\begin{table}
\begin{center}
\caption{The total cross section for the hard subprocess
$g+g\rightarrow B_{cJ,L=1}^*+b+\bar{c}$ (gluon fusion into
$P$-wave excited states $B_{cJ,L=1}^*$) at different C.M.
energies. The input parameters are taken as those used in
Ref.\cite{berezhnoy}: $m_b=5.0GeV$, $m_c=1.7GeV$, $M=6.7GeV$ and
the running $\alpha_s$ is fixed to $0.2$ {\it etc}.}\vspace{4mm}
\begin{tabular}{|c||c|c|c|c|c|c|}
\hline\hline ~C.M. energy (GeV)~ & ~20GeV~ & ~40GeV~ &
 ~60GeV~ & ~80GeV~ & ~100GeV~ & ~200GeV~\\
\hline $\sigma(^1P_1)(pb)$ & 0.367 & 0.743 & 0.657 & 0.538 & 0.439 & 0.195\\
\hline $\sigma(^3P_0)(pb)$ & 0.184 & 0.207 & 0.175 & 0.141 & 0.114 & 0.0496\\
\hline $\sigma(^3P_1)(pb)$ & 0.346  & 0.598 & 0.503 & 0.402 & 0.324 & 0.139\\
\hline $\sigma(^3P_2)(pb)$ & 0.721  & 1.49  & 1.31  & 1.06  & 0.862 & 0.374\\
\hline\hline
\end{tabular}
\label{tabsub}
\end{center}
\end{table}

\begin{figure}
\centering
\includegraphics[width=0.50\textwidth]{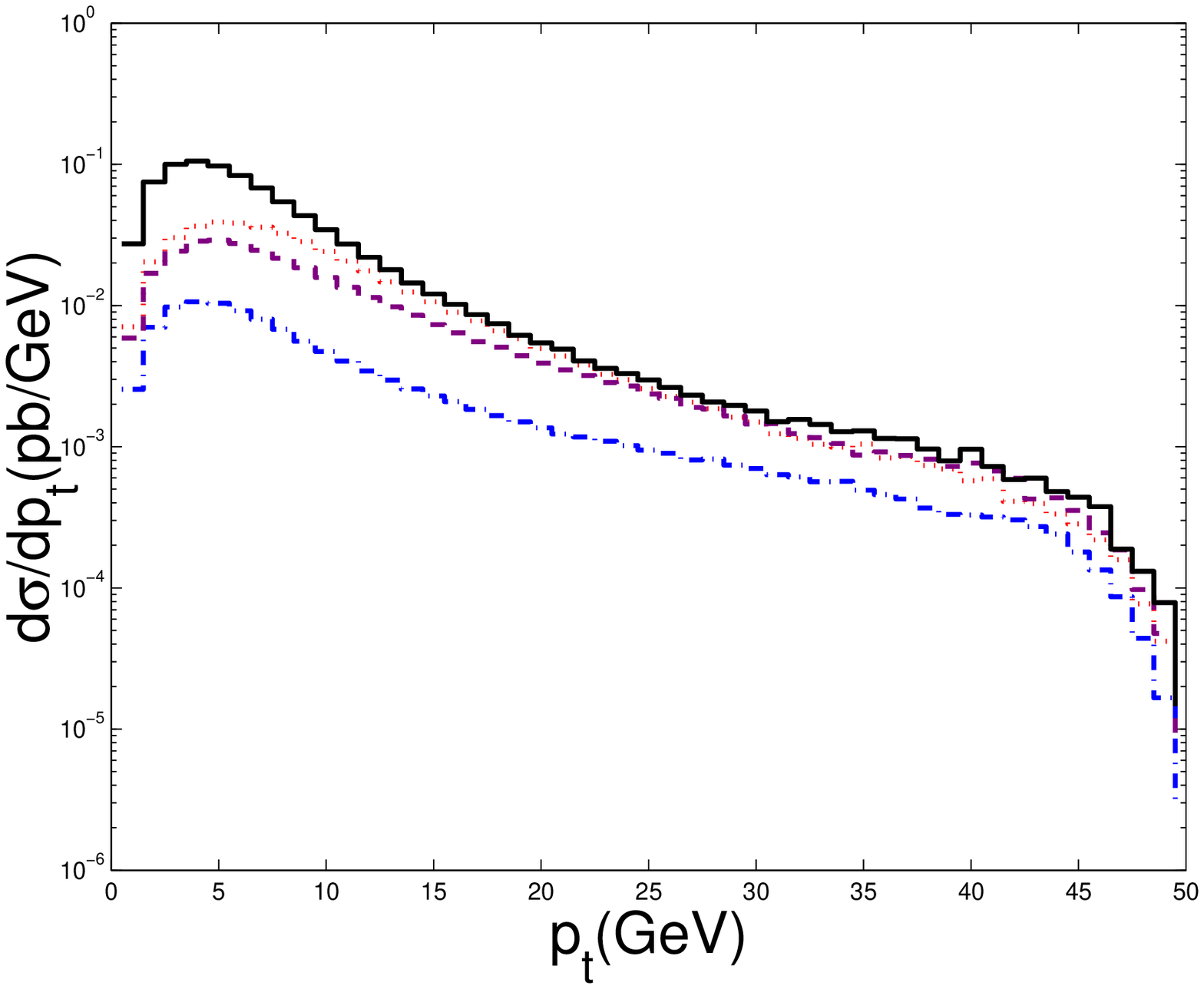}%
\includegraphics[width=0.50\textwidth]{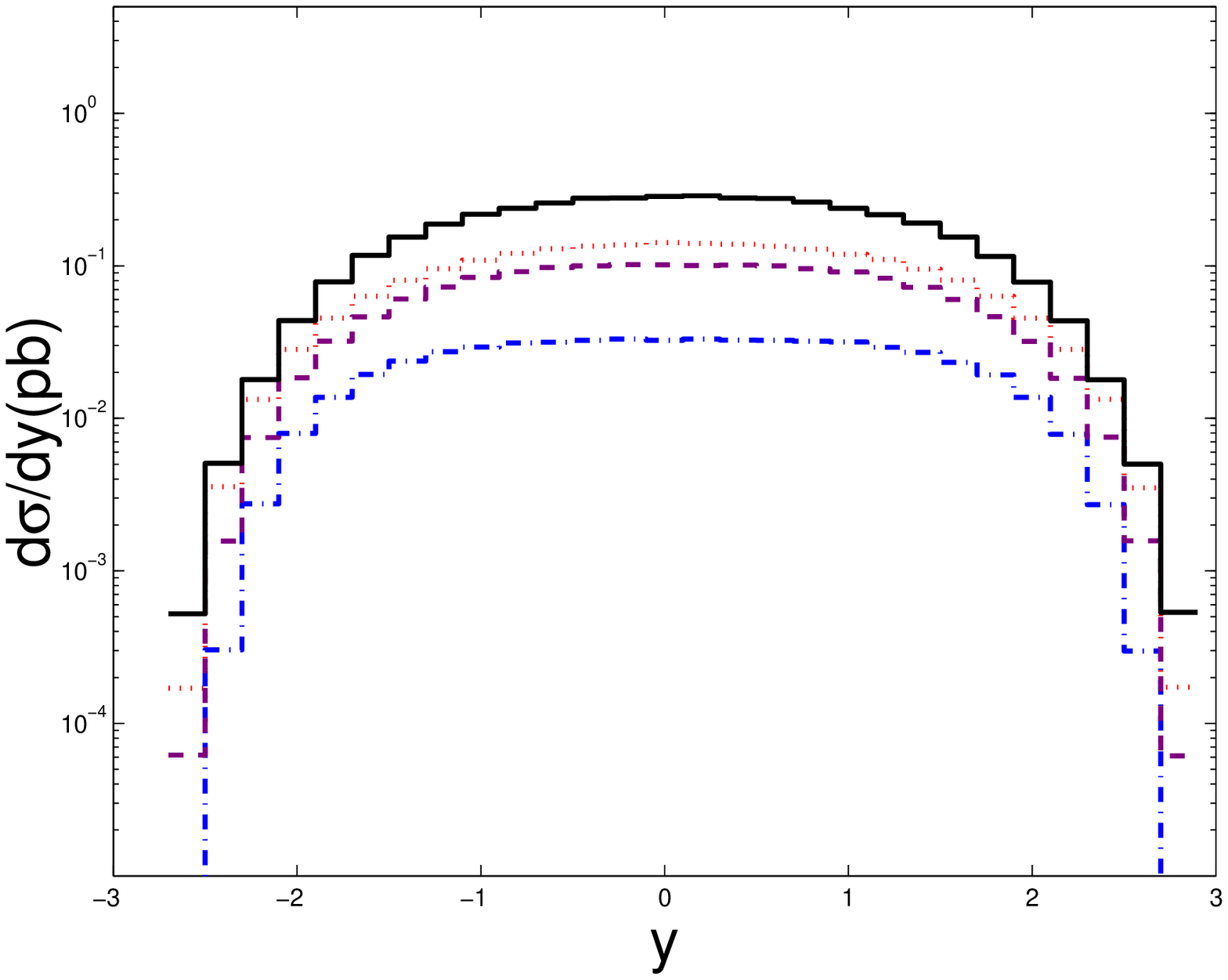}
\caption{$p_t$ and $y$ differential distributions for the
subprocess, where $c\bar{b}$ is in $P$-wave states. The dotted
line, the dash-dot line, the dash line and the solid line are for
$^1P_1$, $^3P_0$, $^3P_1$ and $^3P_2$, respectively.}
\label{subwavep}
\end{figure}

For further checks, we have performed several test runs. With the
option ISUBONLY=$1$, we obtained the transverse momentum
$p_\mathrm{t}$ and rapidity $y$ distributions and the total
cross-sections for the sub-process $gg\rightarrow
B_{cJ,L=1}^*+\bar{c}+b$. The results by taking the same parameters
as those used in Refs.\cite{berezhnoy,berezhnoy2} are shown in the
TABLE.\ref{tabsub} and in FIG.\ref{subwavep}. One may see that the
results for the total cross-section of the hard subprocess
$g+g\rightarrow B_{cJ,L=1}^*+\bar{c}+b$ with various C.M. energies
are agree with those in Ref.\cite{berezhnoy} (TABLE.I there)
within the MC error, and the $p_t$ and $y$ distributions are
consistent well with the ones obtained in
Refs.\cite{berezhnoy,berezhnoy2}. Since our present method is
quite different from that in Refs.\cite{berezhnoy,berezhnoy2}, the
comparison is really a good check for the generator and also for
the results in Refs.\cite{berezhnoy,berezhnoy2}. In fact, in
Refs.\cite{berezhnoy,berezhnoy2}, the derivative over the relative
momentum $q$ of the constitute quarks in the $B_c$ meson, which is
necessary for $P$-wave production, is done numerically; while in
our case, it is done analytically with the help of the FDC program
\cite{fdc}.

\begin{table}
\caption{Generation parameters used in the sample generation.}
\begin{center}
\begin{tabular}{l}
\hline
.............. INITIAL PARAMETERS ................\\
GET THE RESULT FOR Bc IN $^{3}P_0$\\
Bc IN COLOR-SINGLET STATE\\
GENERATE EVNTS 30000000 FOR TEVA ENERGY(GEV) $0.196E+4$\\
$***************************************$\\
$*\;\;$ USING SUBPROCESS: $g+g\rightarrow Bc+b+\bar{c}$$\;\;\;\;\;*$\\
$***************************************$\\
\noindent M\_\{Bc\}=$6.700$$\;\;$M\_\{B\}=
$5.000$$\;\;$M\_\{C\}=$1.700$
$\;\;$f\_\{Bc\}=$0.2879$ \\
Q2 TYPE= 3 $\;\;\;$ ALPHAS ORDER= LO\\
THE NEW VERSION OF OUTER PDFs WHICH ARE FROM WWW\\
PDF: CTEQ6L; ALPHA IN LO\\
USING PYTHIA IDWTUP= 3\\
PTCUT =0.000$\;\;\;$GeV\\
NO RAPIDITY CUT\\
USING VEGAS: NUMBER IN EACH ITERATION= 200000  ITERATION= 30\\
.............. END OF INITIALIZATION ..............\\
\hline
\end{tabular}
\label{output}
\end{center}
\end{table}

\begin{figure}
\centering
\includegraphics[width=0.50\textwidth]{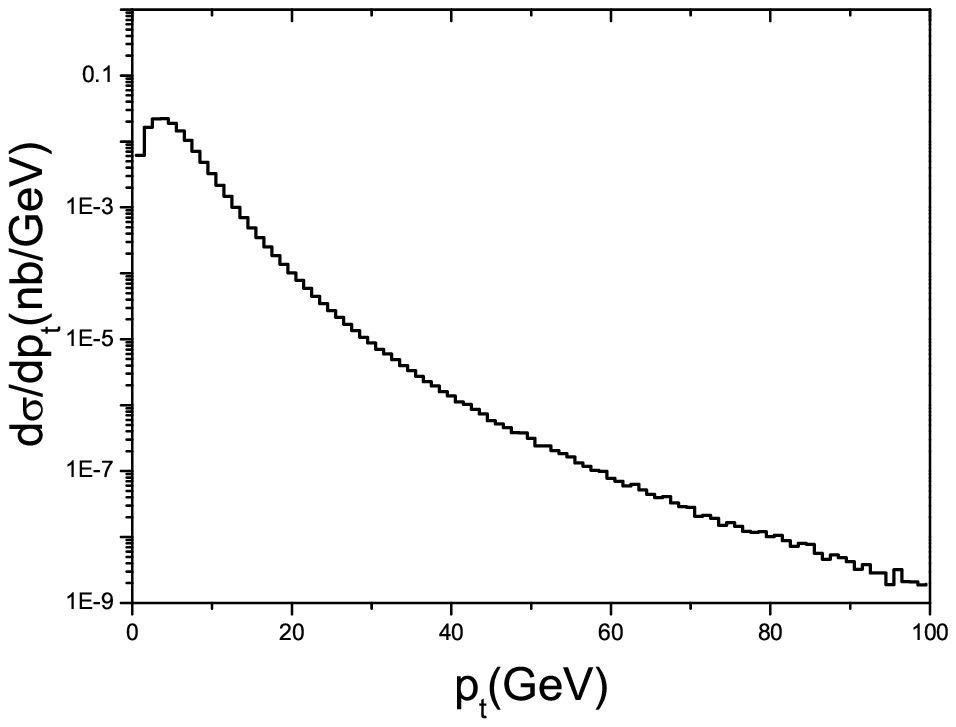}%
\includegraphics[width=0.50\textwidth]{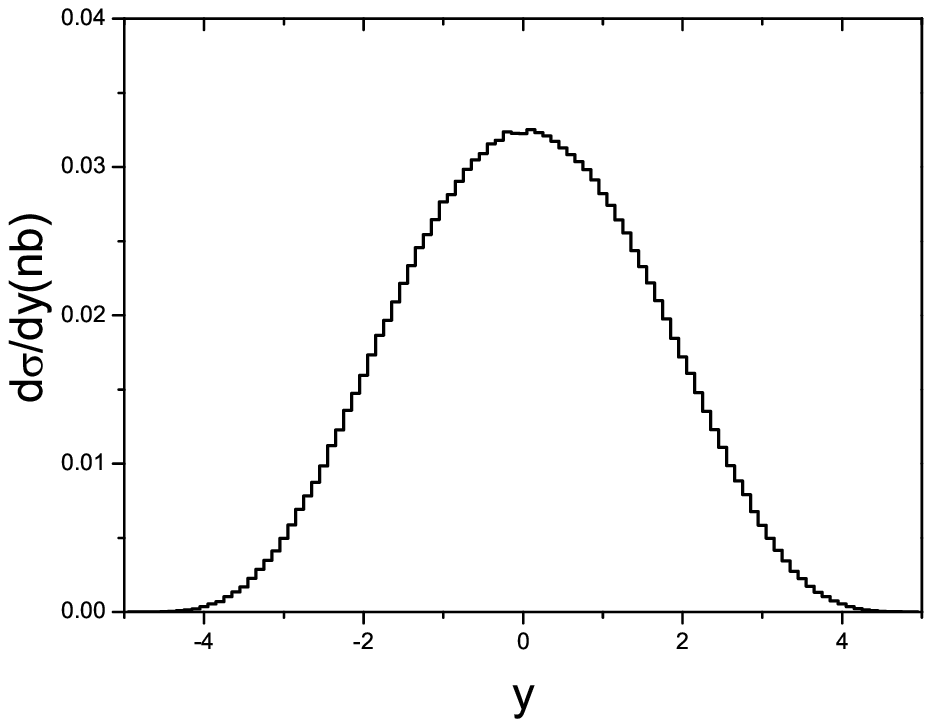}%
\caption{Test run results of the hadronic production of $^3P_0$
state at TEVATRON. The transverse momentum $p_\mathrm{t}$ and
rapidity $y$ distributions for the CTEQ6L by using $\alpha_s$ in
the leading order (LO) and adopting the characteristic energy
scale squared $Q^2=p_{t}^2+M_{(c\bar{b})}^2$.} \label{totpty}
\end{figure}

Finally, we show a test run for the hadronic production of
$(^3P_0)$ $c\bar{b}$-quarkonium at the TEVATRON. When running the
program, the initialization is shown as a screen snap-shot in
TABLE~\ref{output}. Here the value of $f_{Bc}=0.2879$ corresponds
to $R'(0)=0.448 GeV^{5/2}$. The results are shown in
FIG.\ref{totpty}.

\subsection{About VEGAS}

\begin{figure}
\centering
\includegraphics[width=0.65\textwidth]{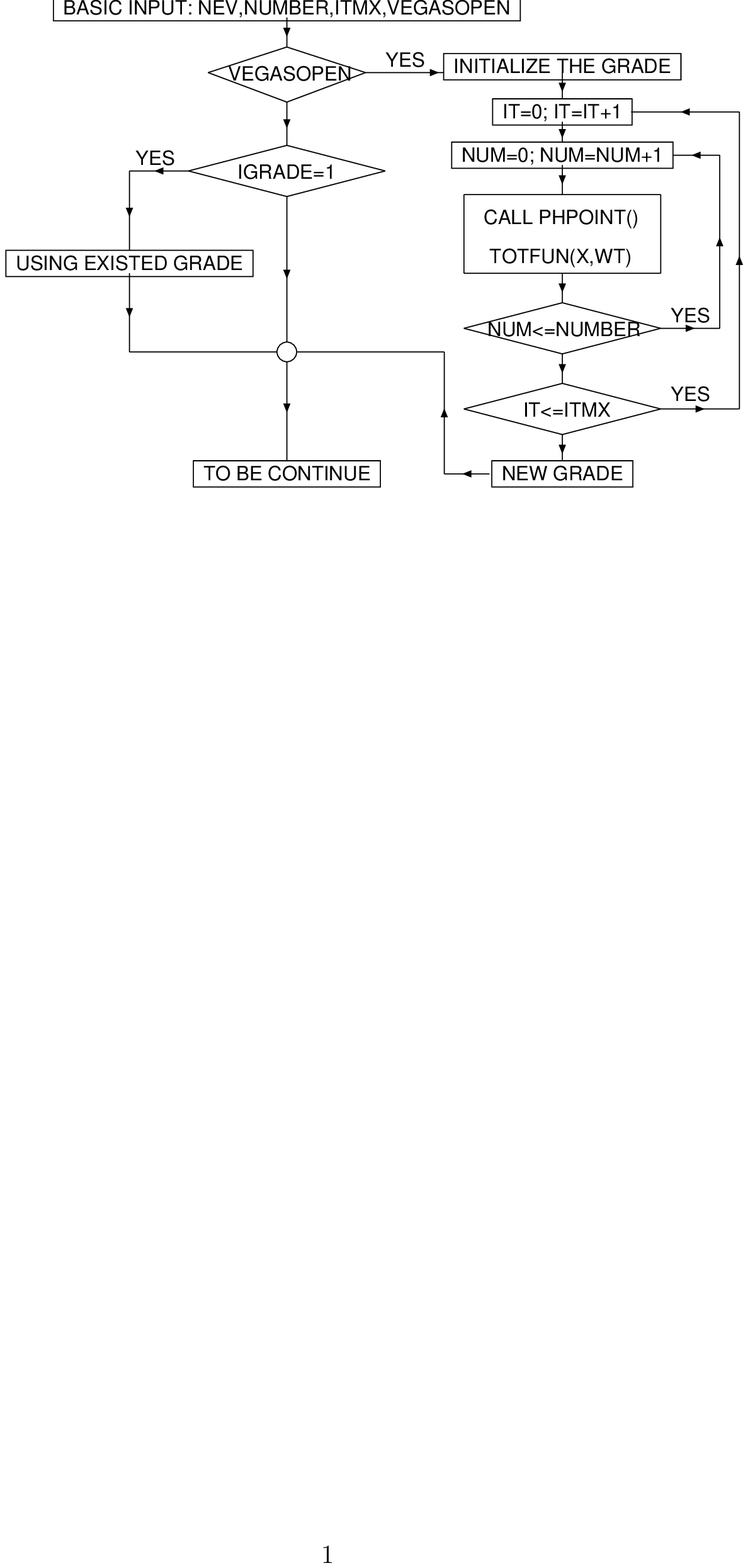}
\caption{Flow chart for using VEGAS. The meaning of the symbols
can be found in the program.} \label{chart}
\end{figure}

This subsection is devoted to give a supplementary explanation on
VEGAS.

By using VEGAS, one can obtain an important sampling-function (or
simply called grade function) into the file grade.dat, and as it
is well known, by using such function to do the event simulation,
the efficiency can be greatly increased \cite{pythia}. A
modification to the original version in Ref.\cite{gpl} is done so
as to record the obtained grade function into the data file
grade.dat. Such small trick is useful, i.e. when one wants to
generate the events with the same input parameters, then one can
directly apply the grade function obtained by the previous running
and does not need to run VEGAS again. The flow chart for this part
is presented in FIG.\ref{chart}, where as mentioned above, one can
use the existent grade function to do the initialization further
and then to obtain a more precise grade function.

In calling VEGAS, we need to input two parameters, i.e. NUMBER
(maximum total number of the times to call the integrand in each
iteration) and ITMX (maximum number of allowed iterations). These
two parameters should be adjusted in a suitable way so as to
obtain a grade function in good precision and also to save the CPU
as much as possible. One subtle point is that, in VEGAS, the
default number of the column bin is 50 (i.e. the region of (0,1)
is divided into 50 pieces); by taking a bigger proper value for
the column bins, one can improve the production efficiency but
need more iterations (or numbers) to obtain a stable result. In
practice, the values for NUMBER, ITMX and the column bins should
be carefully chosen to obtain a best important sampling-function
within the least of time. In our program, we keep the column bins
to be 50 (according to our experience, it is enough); however the
user who wants to do some very precise studies, a proper variation
of column bins might be needed.

\section{Conclusions}

An upgrade version of $B_c$ meson generator BCVEGPY2.0 is given.
In this new version, the hadronic production of $P$-wave states up
to $v^2$, that include both the color-singlet components
$B_{cJ,L=1}^*$ and the color-octet components $|(c\bar b)_{\bf
8}(^{1}S_{0}) g\rangle$ and $|(c\bar b)_{\bf 8}(^{3}S_{1})
g\rangle$, is available. Since almost all of these $P$-wave
excited states decay to the ground state $B_c$ meson through the
electromagnetic transition (if the phase space for strong
interaction transitions is not allowed) finally, thus in fact the
$P$-wave production provides additional contributions to the
hadronic production of the $B_c$ meson. In the new version, the
hadronic production of $S$-wave $c\bar{b}$-quarkonium from the
quark-antiquark annihilation subprocess is also included for
comparison, although the contributions from the quark-antiquark
annihilation mechanism to the $B_c$ hadronic production, in
comparison with those from gluon-gluon fusion mechanism, can be
ignored safely at LHC/TEVATRON energies in the allowed kinematic
region.

To achieve a compact program, we apply the FDC program to
manipulate the amplitude of the dominant gluon-gluon fusion
subprocess for the production corresponding to the four
color-singlet $P$-wave states \cite{cww}. The correctness of the
program is guaranteed by various checks. The generator is
interfaced with PYTHIA, which takes care of producing the full
event and filling the standard PYTHIA event common block. In view
of the prospects for $B_c$ physics at Tevatron and at LHC, the
generator offers a platform for further experimental studies.

\vspace{10mm}
{\bf\Large Acknowledgement:} The work was supported in part by
Nature Science Foundation of China (NSFC). The authors would like
to thank G.M. Chen, S.H. Zhang, A. Belkov, S. Shulga, L. Han, Y.
Jiang and Y.N. Gao for helpful discussions and suggestions. One of
the author (X.G. Wu) would like to thank A.V. Berezhnoy for
discussions on the $P$-wave production.

\appendix
\section{Additional routines and functions for calculating the
hadronic production of $B_{cJ,L=1}^*$}

In this Appendix, additional subroutines and functions for
calculating the amplitudes for the gluon-gluon fusion subprocess,
$gg\rightarrow B_{cJ,L=1}^*+b+\bar{c}$, are explained.

\noindent {\bf SUBROUTINE ampI$\_$1p1(cc)}

\noindent Purpose: with I=$1,2,\cdots,36$ to compute the amplitude
of the 36 Feynman diagrams of the gluon-gluon fusion subprocess
correspondingly, with the $(c\bar{b})$-quarkonium in $^1P_1$
state.

\noindent Real*8 cc=: $cc=\pm
\sqrt{\frac{9R'(0)^2}{2\pi}\frac{1}{72M^3}}$, where $R'(0)^2$ is
the square of the first derivative of the radial wave function at
the origin, and $M$ is the mass of the $(c\bar{b})$-quarkonium.
Whether it takes the plus or the minus sign is our inner
convention for the 36 Feynman diagrams and it depends on how to
decompose the Feynman diagrams.

\noindent{\bf SUBROUTINE ampI$\_$3p0(cc)}

\noindent Purpose: with I=$1,2,\cdots,36$, to compute the
amplitude of the 36 Feynman diagrams of the gluon-gluon fusion
subprocess correspondingly, with the $(c\bar{b})$-quarkonium in
$^3P_0$ state.

\noindent Real*8 cc=: $cc=\pm
\sqrt{\frac{9R'(0)^2}{2\pi}\frac{1}{216M^3}}$, where $R'(0)^2$ is
the square of the first derivative of the radial wave function at
the origin and $M$ is the mass of the $(c\bar{b})$-quarkonium.
Whether it takes the plus or the minus sign is our inner
convention for the 36 Feynman diagrams and it depends on how to
decompose the Feynman diagrams.

\noindent {\bf SUBROUTINE ampI$\_$3p1(cc)}

\noindent Purpose: with I=$1,2,\cdots,36$, to compute the
amplitude of the 36 Feynman diagrams of the gluon-gluon fusion
subprocess correspondingly, with $(c\bar{b})$-quarkonium in
$^3P_1$ state.

\noindent Real*8 cc=: $cc=\pm
\sqrt{\frac{9R'(0)^2}{2\pi}\frac{1}{144M^3}}$, where $R'(0)^2$ is
the square of the first derivative of the radial wave function at
the origin and $M$ is the mass of the $(c\bar{b})$-quarkonium.
Whether it takes the plus or the minus sign is our inner
convention for the 36 Feynman diagrams and it depends on how to
decompose the Feynman diagrams.

\noindent {\bf SUBROUTINE ampI$\_$3p2(cc)}

\noindent Purpose: with I=$1,2,\cdots,36$, to compute the
amplitude of the 36 Feynman diagrams of the gluon-gluon fusion
subprocess correspondingly, with $(c\bar{b})$-quarkonium in
$^3P_2$ state.

\noindent Real*8 cc=: $cc=\pm
\sqrt{\frac{9R'(0)^2}{2\pi}\frac{1}{72M^3}}$, where $R'(0)^2$ is
the square of the first derivative of the radial wave function at
the origin and $M$ is the mass of the $(c\bar{b})$-quarkonium.
Whether it takes the plus or the minus sign is our inner
convention for the 36 Feynman diagrams and it depends on how to
decompose the Feynman diagrams.

\noindent {\bf FUNCTION amps20$\_$1p1()}

\noindent Purpose: to compute the square of the whole amplitude
for a particular combination of two gluons' polarization states
and the polarization vector of $^1P_1$ $c\bar{b}$-quarkonium
state.

\noindent {\bf FUNCTION amps2$\_$1p1()}

\noindent Purpose: to compute the square of the whole amplitude
with the help of the FUNCTION amps20$\_$1p1(), where the summation
over the gluons' polarization states and the polarization vector
of $^1P_1$ $c\bar{b}$-quarkonium state is done.

\noindent {\bf FUNCTION amps20$\_$3p0()}

\noindent Purpose: to compute the square of the whole amplitude
for the production of $^3P_0$ $c\bar{b}$-quarkonium and for a
particular combination of the two gluons' polarization states.

\noindent {\bf FUNCTION amps2$\_$3p0()}

\noindent Purpose: to compute the square of the whole amplitude
with the help of the FUNCTION amps20$\_$3p0(), where the summation
over the gluons' polarization states is done.

\noindent {\bf FUNCTION amps20$\_$3p1()}

\noindent Purpose: to compute the square of the whole amplitude
for a particular combination of two gluons' polarization states
and the polarization vector of $^3P_1$ $c\bar{b}$-quarkonium
state.

\noindent {\bf FUNCTION amps2$\_$3p1()}

\noindent Purpose: to compute the square of the whole amplitude
with the help of the FUNCTION amps20$\_$3p1(), where the summation
over the gluons' polarization states and the polarization vector
of $^3P_1$ $c\bar{b}$-quarkonium state is done.

\noindent {\bf FUNCTION amps20$\_$3p2()}

\noindent Purpose: to compute the square of the whole amplitude
for a particular combination of two gluons' polarization states
and the polarization tensor of $^3P_2$ $c\bar{b}$-quarkonium
state.

\noindent {\bf FUNCTION amps2$\_$3p2()}

\noindent Purpose: to compute the square of the whole amplitude
with the help of the FUNCTION amps20$\_$3p2(), where the summation
over the gluons' polarization states and the polarization vector
of $^3P_2$ $c\bar{b}$-quarkonium state is done.

\noindent {\bf SUBROUTINE genpolar3(p, ep, pm)}

\noindent Purpose: to generate the polarization vectors (ep) of a
vector meson with mass pm and momentum p.

\noindent real*8 p(4), ep(4,2:4): p(4) is the momentum for the
vector meson, $p(1)=E$, $p(2)=p_x$, $p(3)=p_y$ and $p(4)=p_z$.
ep(4,2:4) records three polarization vectors of the vector meson,
each of which has four components. The explicit formulas of $ep$
can be found in appendix A of Ref.\cite{cww}.

\noindent {\bf SUBROUTINE genpolar(p, ep)}

\noindent Purpose: to generate the polarization vectors (ep) of a
gluon with momentum p.

\noindent real*8 p(4), ep(4,2:3): ep(4,2:3) records two transverse
polarization vectors of the gluon.

\noindent {\bf SUBROUTINE gentensor(p, ep, pm)}

\noindent Purpose: to generate the polarization tensor (ep) for
the meson in $^3P_2$ state with mass pm and momentum p.

\noindent real*8 p(4), ep(4,2:6): ep(4,2:6) records five
polarization tensors of the meson in $^3P_2$ state. The explicit
formulas of $ep$ can be found in appendix A of Ref.\cite{cww}.

There are also some accessory subroutines and functions that are
used to simplify the amplitude and to make the amplitude more
compact and not difficult to be understood, such as the {\bf
SUBROUTINE genppp} which is used to generate short notations for
all the typical linear combination of dotted products of two
momenta, thus for shortening the paper we do not explain them
precisely here.

\end{document}